\documentclass[prl, twocolumn,superscriptaddress,nofootinbib, amsmath,amssymb, aps,a4paper, accepted=2021-03-14]{quantumarticle}
\pdfoutput=1
\usepackage[utf8]{inputenc}
\usepackage[T1]{fontenc}
\usepackage[UKenglish]{babel}
\usepackage{graphicx}
\usepackage{amsmath,amssymb,amsthm}
\usepackage{color}
\usepackage{psfrag}
\usepackage{ifsym}
\usepackage{epstopdf}
\usepackage{dsfont}
\usepackage{multirow} 
\usepackage{paralist}
\usepackage{enumitem}
\usepackage[normalem]{ulem}
\usepackage{units}
\usepackage{tabularx}
\usepackage{bbm}
\usepackage[utf8]{inputenc}
\usepackage[T1]{fontenc}
\usepackage[colorlinks]{hyperref}
\usepackage{xspace}
\usepackage{mathtools}
\usepackage{dsfont}
\usepackage{lipsum}
\usepackage{siunitx}
\usepackage{pgf,tikz}
\usepackage{pgfplots}
\usepackage{footnote}
\usepackage{soul,xcolor}
\setstcolor{blue}
\usepackage{subfigure}
\usepackage{natbib}
\usepackage{hyperref}
\usepackage{chapterbib}
\usepackage[numbers]{natbib}
\usepackage{bibunits}

\DeclareMathOperator{\Tr}{Tr}
\DeclareMathOperator{\tr}{tr}

\newcommand{\bra}[1] {\langle #1 |}
\newcommand{\ket}[1] {| #1 \rangle}

\newcommand{\ketbra}[2]{ | #1 \rangle\!\langle #2 |}
\newcommand{\proj}[1]{ | #1 \rangle\!\langle #1 |}

\newcommand{\id}{\mathds{1}}

\newcommand{\Ket}[1] {| #1 \rangle\!\rangle}

\newcommand{\KetBra}[2]{ | #1 \rangle\!\rangle\!\langle\!\langle #2 |}

\newcommand{\ba}{\begin{eqnarray}}
\newcommand{\ea}{\end{eqnarray}}
\newcommand{\be}{\begin{equation}}
\newcommand{\ee}{\end{equation}}
\newcommand{\bs}{\begin{split}}
\newcommand{\es}{\end{split}}
\newcommand{\bef}{\begin{figure}[!h]}
\newcommand{\eef}{\end{figure}}
\newcommand{\etal}{\emph{et al.}\@\xspace}
\newcommand{\trho}{\tilde{\rho}}

\newtheorem*{lem*}{Lemma}

\hypersetup{pdftitle={{Witnessing quantum non-Markovianity}}}

\begin{document}


\title{Witnessing quantum memory in non-Markovian processes}

\author{Christina Giarmatzi}
\email{christina.giar@gmail.com}
\affiliation{Centre for Engineered Quantum Systems, School of Mathematics and Physics, University of Queensland, QLD 4072 Australia}
\affiliation{University of Technology Sydney, Centre for Quantum Software and Information, Ultimo NSW 2007, Australia}
\author{Fabio Costa}
\affiliation{Centre for Engineered Quantum Systems, School of Mathematics and Physics, University of Queensland, QLD 4072 Australia}

\begin{bibunit}[apsrev4-1_modified]


\maketitle

\textbf{We present a method to detect quantum memory in a non-Markovian process. We call a process Markovian when the environment does not provide a memory that retains correlations across different system-environment interactions. We define two types of non-Markovian
processes, depending on the required memory being classical or quantum. We formalise this distinction using the process matrix formalism, through which a process is represented as a multipartite
state. Within this formalism, a test for entanglement in a state can be mapped to a test for quantum memory in the corresponding process. This allows us to apply separability criteria and entanglement witnesses to the detection of quantum memory. We demonstrate the method in a simple model where both system and environment are single interacting qubits and map the parameters that lead to quantum memory.
As with entanglement witnesses, our method of witnessing quantum memory provides a versatile experimental tool for open quantum systems.}

\section{Introduction}
In any quantum device, the system that carries the information unavoidably interacts with its environment introducing noise. Studying the dynamics of such system-environment interactions is the field of \emph{open quantum systems}~\cite{Breuer} and it is nowadays more relevant than ever. {As quantum devices begin to demonstrate an advantage over classical ones~\cite{Arute2019}, they increasingly rely on} Noisy Intermediate-Scale Quantum (NISQ) technology, whose main challenge is noise~\cite{Preskill2018}.

{Noise models typically rely on the assumption of \emph{Markovianity}, i.e., that the environment does not keep memory of past interactions with the system. However, this assumption typically fails in realistic scenarios, as information stored in the environment can keep track of past interactions with the system and affect its future dynamics.} For example, this was demonstrated to occur in the IBM quantum computing platform~\cite{morris2019}. In the study of such memory effects, an important distinction is whether the memory can be represented classically or {requires genuinely quantum degrees of freedom.}
{The two scenarios can lead to radically different noise models and strategies to compensate it. {For example, in the presence of classical memory one could monitor the system and correct its evolution through an appropriate classical feedback. However, this method
fails for quantum temporal correlations mediated by the environment.} It is therefore desirable to find efficient methods to discriminate quantum vs classical memory.} 

Most models of open quantum systems with memory regard the process as a dynamical map which maps the system from one time-step to the other~\cite{Piilo08, Wolf2008, Breuer09, Rivas10, Hou11, Chruscinski14, Rivas2014,Breuer16, LI20181,Vega:2017aa}. Within this approach, many models of processes with classical memory have been developed~\cite{Shapiro1987, Caves1987, wiseman1993, budini2001, Zhou2010, BORDONE2012, Bodor2013, Xu2013, Vacchini2013, Budini2018}. However, dynamical maps study only two-time correlations (time of input and output of the map) and multi-time-correlations cannot be fully captured. {Furthermore, dynamical maps are in general ill-defined in the presence of initial system-environment correlations \cite{Shaji2005, Pechukas1994, Stelmachovic2001, Schmid2019}, although such correlations can be responsible for non-Markovianity.}

Here, {we introduce a definition of quantum process with classical memory based on} an approach that captures multi-time correlations, {originally introduced by Lindblad~\cite{Lindblad:1979aa} and Accardi \textit{et al.}~\cite{Accardi82}, and recently reformulated within the comb formalism~\cite{chiribella09b} by Pollock \textit{et al.}~\cite{Pollock:2018aa}, Fig.~\ref{fig:1}.} {We} provide a technique to {efficiently} detect {the presence of} quantum memory {in a non-Markovian process,} without {requiring} full tomography. We use the process matrix formalism~\cite{Oreshkov12,Oreshkov16} to write the process as a multipartite state. For a specific partition of the state, {classical memory implies separability, while} entanglement proves quantum memory. Therefore, we can employ all the known techniques that verify entanglement and use them to prove non-Markovianity with quantum memory. 


\bef
\center
\includegraphics[width=.8\linewidth]{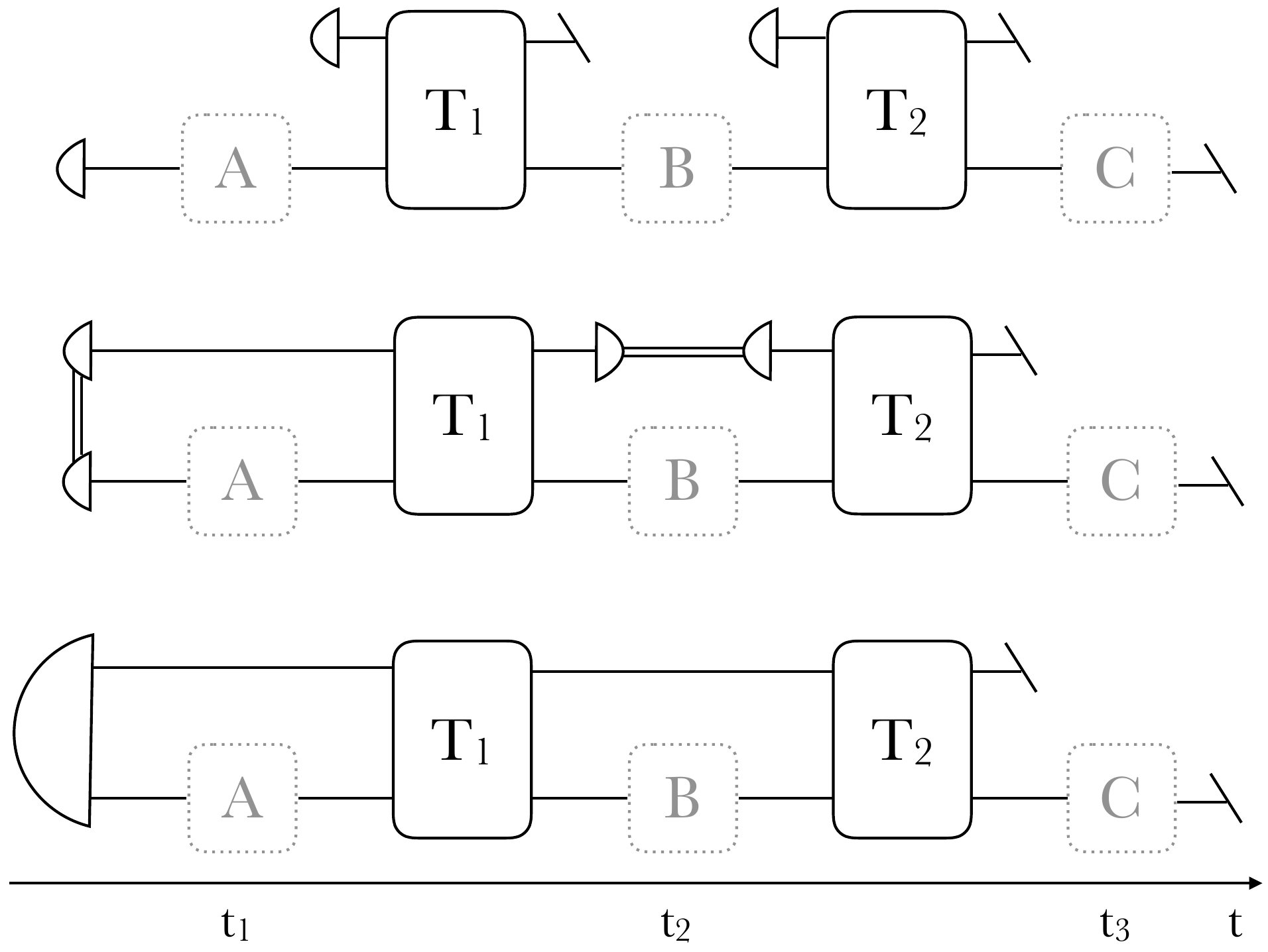}
\caption[]{Three types of processes with three time-steps: Markovian (top), where the environment has no memory, classical memory (middle), where classical information from the system is carried by the environment, and quantum memory (bottom), where there are initial quantum correlations that travel across the process. $A$, $B$ and $C$ are measurement stations where arbitrary quantum operations can be performed.}
\label{fig:1}
\eef

To illustrate our method of detecting quantum memory, we use entanglement witnesses to obtain \emph{witnesses for quantum memory} for the following toy model: system and environment are qubits jointly prepared in a maximally entangled state and later interact according to the Ising model, in between two measurement stations $A$ and $B$ for the system. A quantum memory witness corresponds to a set of operations at $A$ and $B$. As separability criteria for the search of witnesses we use the positive partial transpose (PPT) applied on the state~\cite{Peres:1996aa} and on symmetric extensions of the state~\cite{Doherty:2004aa}. {To find a witness, we cast each criterion as a SemiDefinite Program that can be solved efficiently~\cite{nesterov87}. This also allows us to restrict the search for witnesses, possibly tailored to experimental limitations.}

\section{Quantum processes}
We use the process matrix formalism~\cite{Oreshkov12,Oreshkov16} to describe quantum processes. {{We consider a system that evolves through a given sequence of time steps. Each time step can be seen as a ``measurement station'', labelled $A, B,\dots$, where an experimenter can intervene on the system, e.g.\ by measuring or transforming it.}
 Each operation can be represented by a completely positive map (CP) $\mathcal{M}^{A_I\rightarrow A_O}$ that maps the input system $A_I$ to the output system $A_O$ of the operation.  {A general intervention on the system is described by a \emph{quantum instrument}~\cite{davies70}: a collection of CP maps} ${\cal J}^A = \{ {\cal M}^A\}$, corresponding to all possible outcomes of a measurement, with ${\sum}_{{\cal M}^A\in {\cal J}^A} {\cal M}^A$ being a CP and trace preserving (CPTP) map. } {The joint probability for observing the outcomes corresponding to maps ${\cal M}^A, {\cal M}^B,\dots$ at the respective measuring stations is given by}
\ba
p({\cal M}^{A}, {\cal M}^{B},\cdots | {\cal J}^{A}, {\cal J}^{B}, \cdots) = \nonumber \\ 
\Tr[W^{A_IA_OB_IB_O\cdots} (M^{A_IA_O} \otimes M^{B_IB_O} \otimes \cdots)],
\label{eq:prob}
\ea
where $M^{A_IA_O}, M^{B_IB_O}, \cdots$ are the Choi matrices~\cite{jamio72, Choi1975} of the maps ${\cal M}^{A}, {\cal M}^{B}, \dots$ respectively, and $W^{A_IA_OB_IB_O\cdots}$ is the \emph{process matrix} that lives on the combined Hilbert spaces of all input and output systems. The Choi matrix $M^{A_IA_O}\in {\cal L}({\cal H}^{A_I}\otimes{\cal H}^{A_O})$, isomorphic to a CP map ${\cal M}^{A} : {\cal L}({\cal H}^{A_I}) \rightarrow {\cal L}({\cal H}^{A_O})$ is defined as $M^{A_IA_O} := [{\cal I} \otimes {\cal M}(\KetBra{\id}{\id})]^T$, where $\cal I$ is the identity map, $\Ket{\id} = \sum_{j=1}^{d_{A_I}}\ket{jj} \in {\cal H}^{A_I} \otimes {\cal H}^{A_I}$, $\{\ket{j}\}^{d_{A_I}}_{j=1}$ is an orthonormal basis on ${\cal H}^{A_I}$ and ${T}$ denotes matrix transposition in that basis and some basis of ${\cal H}^{A_O}$.
{For the scenarios relevant here, where events are in a fixed causal order, process matrices are identical to quantum combs~\cite{chiribella09b} and process tensors~\cite{Pollock:2018aa}. In turn, these objects correspond to the Choi matrix of a quantum channel with memory~\cite{Kretschmann2005}.} 

In this language, it was found that the process matrix of a Markovian process{, where the state of the environment is reset before each time step (Fig.~\ref{fig:1}, top)}, must have the following form~\cite{costa2016, Giarmatzi:2018aa,Pollock:2018ab},
\ba
W^{AB\cdots}_{\textup {M}} = \rho^{A_I}\otimes{T}^{A_OB_I}\otimes \cdots,
\label{eq:Wab_m}
\ea
where {$\rho$ is the density matrix of the system before the first time step and} ${T}^{A_OB_I}$ is the Choi matrix of a channel $\cal T^{A\rightarrow B}$, defined as above but without the transposition. {This means that a Markovian process can be written as a tensor product of the initial state and the Choi matrices of the different channels that connect the measurement stations.} {As we will see later, this approach to non-Markovianity takes into account all possible temporal correlations within a quantum process---unlike any other definitions based on dynamical maps which only capture two-time-step correlations.}

A non-Markovian process with classical memory is one where, during each system-environment interaction, the environment obtains some classical information about the system, which can affect future such interactions {(Fig.~\ref{fig:1}, middle). In other words, the environment can be simulated as a feedback mechanism, which measures the system at each time step and, conditioned on the outcome, affects the system's evolution at future times~\cite{Shapiro1987, Caves1987, wiseman1993}.} {To formalise this definition we describe the evolution between two time steps as a \emph{quantum instrument}: a collection of CP maps that sum up to a CPTP map. Each instrument can depend on the classical information stored in the environment up to that point, namely on all the previous measurement outcomes. We provide a more detailed definition of processes with classical memory in Appendix A1, together with a characterisation of their process matrices.}

{The main result, relevant for what follows, is that a process matrix with classical memory is proportional to a separable state. Indeed, we prove in Appendix A1 that a necessary condition for a process matrix with classical memory is that it has the form}
\ba
W^{AB\cdots}_{\textup{Cl}} = \sum_j \rho^{A_I}_j\otimes T^{A_OB_I}_j \otimes \cdots,
\label{eq:Wab_cl}
\ea
where $\rho_j$, $T_j$ are positive semidefinite matrices.

A \emph{process with quantum memory}, on the other hand, is one that {does not fall within either of the above categories} (Markovian or with classical memory) 
{In such a process (Fig.~\ref{fig:1}, bottom), the environment preserves coherence across {different} interactions with the system, in a way that cannot be simulated by a measurement and feedback model. }

\section{Detecting quantum memory}
From Eq.~\ref{eq:Wab_m} we observe that a Markovian process is a tensor product of a state with the Choi matrix of the channels involved. In what follows we focus on scenarios with two time-steps where operations $A$ and $B$ can be performed. The output of $B$ can be ignored since the only contribution to non-Markovianity stems from the initial system-environment correlations and their later interaction. If such a process has a classical memory, it can be written as $W^{A_IA_OB_I}_{\textup{Cl}} = \sum_j \rho^{A_I}_j\otimes T^{A_OB_I}_j$. Choi's theorem~\cite{Choi1975} ensures that the matrices $T^{A_OB_I}_j$ are positive semidefinite; hence the process matrix can be viewed as a product state, up to normalization. Therefore, $W^{A_IA_OB_I}_{\textup{Cl}}$ can be written as a bipartite separable state $\rho_{sep}^{AB}$, where $A = A_I$, $B = A_OB_I$.

We can now see that detecting entanglement of the state translates to detecting a quantum memory for the process{---if the process cannot be written as a separable state, it cannot be Markovian or with classical memory. Note that quantum memory corresponds to entanglement between $A$ and $B$, which is $A_I$ and $A_OB_I$. This is our main result; we can use known methods of entanglement detection to verify that the environment of a non-Markovian process provides a quantum memory---a feature that has not been explored before. {Previous works have connected entanglement with non-Markovianity---for example in~\cite{Rivas10, Koodynski:2020aa, Chruscinski:2014aa}---however, these approaches are based on dynamical maps and therefore, as we discussed, they are insufficient for a full characterisation of non-markovianity}. We proceed with the detection method, and discuss later possible applications.}

One approach to detecting entanglement is through entanglement witnesses. An entanglement witness is a hermitian operator $Z$ whose expectation value is positive for all separable states, $\langle Z \rangle = \Tr(Z\rho_{sep}) \geq 0.$
A negative expectation value guarantees entanglement of the state, whereas a positive value provides no guarantee for separability. In the case of processes, we define a \emph{quantum memory witness} to be a hermitian operator $Z$ whose expectation value is positive for all process matrices with classical memory
\ba
\langle Z \rangle = \Tr(Z W_{\textup{Cl}}) \geq 0.
\ea

We note that an entanglement witness is also a quantum memory witness for the associated process, but the reverse is not necessarily true. {This has to do with the fact that Eq~\ref{eq:Wab_cl} is a necessary condition for a process with classical memory but not sufficient. That is, not any separable state represents a comb. This is because the set of states is larger than the set of combs; see Fig~\ref{fig:sets}. Combs~\cite{chiribella09b} obey
further linear constraints $W = L(W)$, where $L$ is a projector on the subspace characterised{, e.g.,} in~\cite{araujo15}. Hence, $\Tr(Z W) = \Tr(Z L(W)) = \Tr(L(Z) W)$---$L$ is self adjoint.This means that, if L(Z) is an entanglement witness, then Z is also a witness for classical memory, although not necessarily an entanglement witness. Finally, we conjecture that the space of processes with classical memory is strictly smaller than the intersection of combs and separable states. These are in direct analogy to the fact that not all separable channels can be realised through local operations and classical communication~\cite{Rains1997, Vedral1998, Bennett1999}.} {See Appendix A1 for further discussion.}

\bef
\center
\includegraphics[width=0.95\linewidth]{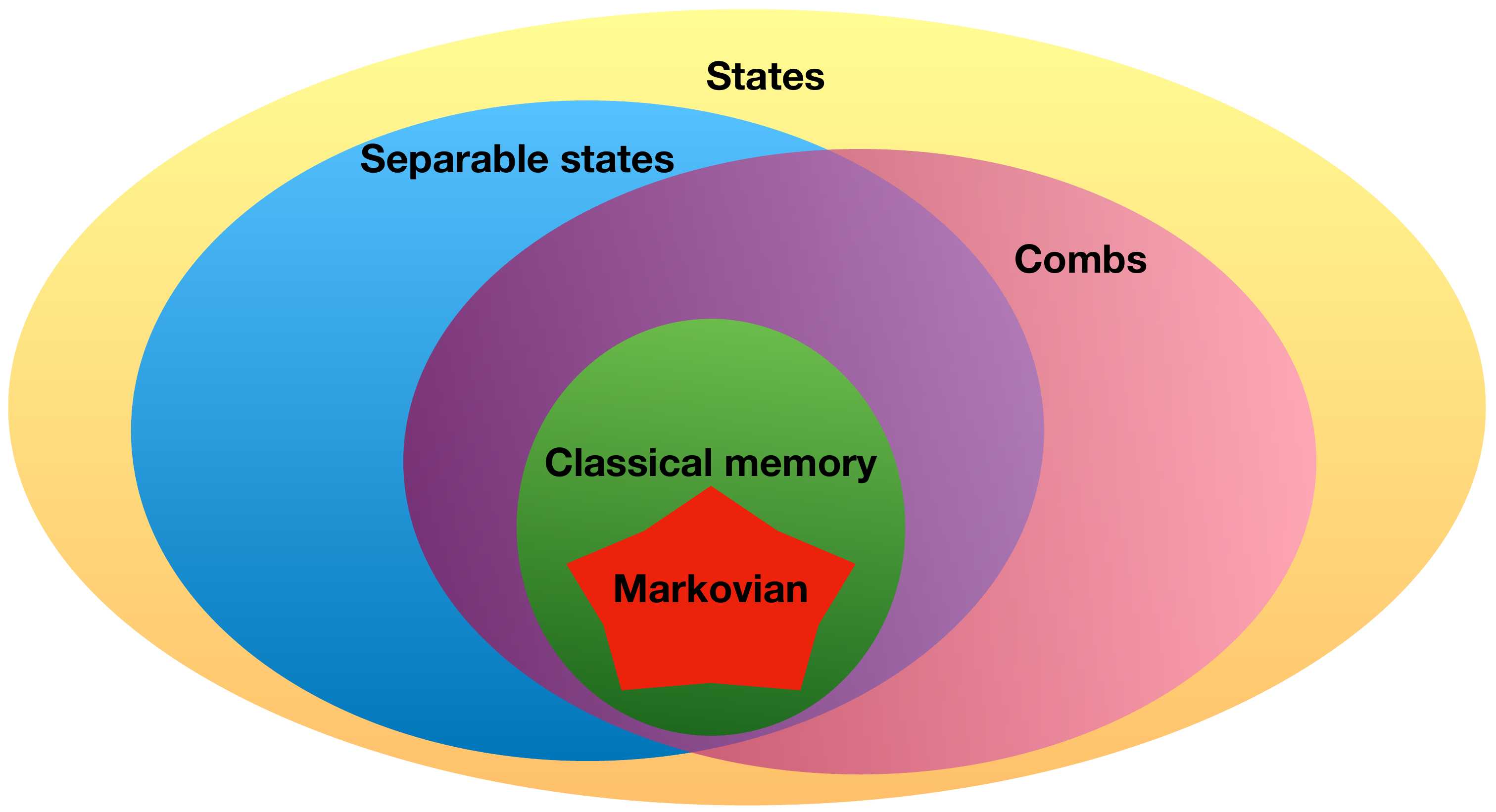}
\caption[]{The space of combs is smaller than the space of states. In turn, the space of combs with classical memory is smaller than that of separable states because combs need to obey further constraints. We conjecture that the space of combs with classical memory is smaller than the intersection of combs and separable states. (During the revision stage of this manuscript, an example confirming the conjecture was presented in~\cite{nery2021simple}.)}
\label{fig:sets}
\eef

A quantum memory witness is a hermitian matrix, hence it can always be decomposed into combinations of CP maps~\cite{araujo15}. From Equation~\ref{eq:prob} we see that the witness corresponds to the different maps of the measurement stations; for example for a bipartite process,
\ba \label{CPdecomposed}
Z = \sum_{i,j} \alpha_{i,j} M^{A_IA_O}_i \otimes M^{B_I}_j.
\ea 
{Crucially, measuring the witness only requires performing the CP maps in \eqref{CPdecomposed} and does not require full process tomography. This makes the technique suitable for experiments, similarly to witnesses for indefinite causal order \cite{Rubinoe1602589, Goswami2018}.}

{We note that detecting a quantum memory in an experimental setup is a procedure similar to that of detecting entanglement. First we need a good guess of the realised process matrix. We proceed by finding a witness of quantum memory, that is, an operator consisting of general quantum operations within the process. Experimentally, a negative expectation value of this operator detects quantum memory.}

{Given the several approaches to non-Markovianity, based on dynamical maps rather than on process matrices, one might wonder if these could also provide a way to witness quantum memory. As we prove in Appendix A2, this is not the case.}

\subsection{Separability criteria}An entanglement witness is obtained through various separability criteria: a property that is proved to hold for all separable states. These criteria provide necessary (but not sufficient) conditions for a state to be separable. The most notable criterion is based on partial transposition: 
if a state is separable, it must have a positive partial transpose (PPT)~\cite{Peres:1996aa} with respect to any subsystem. For example, for a bipartite state $\rho^{AB}$ it is separable if (but not only if) $\rho^{T_A} \geq0$
where $\rho^{T_A}$ denotes partial transposition with respect to subsystem $A$ ($\rho^{T_B}\geq0$ is an equivalent condition, but is superfluous since positivity of $\rho$ and $\rho^{T_A}$ implies positivity of $\rho^{T_B}$). An entanglement witness is obtained through the eigenvector corresponding to the negative eigenvalue of $\rho^{T_A}$. If {$\rho^{T_A}\ngeq0$}, there exists $\ket{\psi}$ such that $\bra{\psi}\rho^{T_A}\ket{\psi} <0$. Then $Z = \ketbra{\psi}{\psi}^{T_A}$ is an entanglement witness as its expectation value is negative for the state $\rho$ and positive for any separable state~\cite{Peres:1996aa}
\ba
\Tr (\ketbra{\psi}{\psi} \rho^{T_A}) <0 \Rightarrow \Tr(Z \rho) <0, \ \textup{and}\nonumber \\
\Tr (Z \rho^A\otimes \rho^B ) = \Tr (\ketbra{\psi}{\psi} \rho_{sep}^{T^A}) \geq 0.
\ea

A different family of separability criteria was introduced by Doherty \emph{et al.}~\cite{Doherty:2004aa}. These criteria are based on the PPT criterion applied on symmetric extensions of the state. For example, for $\rho^{AB}$, one can apply the PPT criterion to the extended state $\tilde{\rho}^{ABA}$ and obtain the following necessary conditions for the state to be separable:
\ba
\tilde{\rho}\geq0,\ \ \tilde{\rho}^{T_A}\geq0,\ \ \tilde{\rho}^{T_B}\geq0,
\ea

One can keep extending the state (i.e. $\rho^{ABAA}$) and obtain new families of conditions, each at least as strong as the previous ones. This creates a hierarchy of families of separability criteria which was proven to be complete---for an entangled state it is guaranteed that some level of the hierarchy will prove non-separability. The first step of the hierarchy is the PPT criterion, the first symmetric extension is the second step, and so on. As we discussed earlier, these results directly extend to define a family of \emph{criteria for classical memory}. 

\subsection{Example}{We use the above separability criteria to detect a quantum memory on a two time-step process, like the one on Fig~\ref{fig:1} (bottom), up until time $t_2$. When the initial system-environment state is a product or separable state, the process is Markovian or with classical memory, respectively, for any unitary $U$, as we prove in the Appendix A3. Hence, initial entanglement is necessary for the environment to provide quantum memory. However, it is not sufficient{, as $T_1$ must also provide an appropriate interaction between system and environment.}}

{We choose the following: system and environment are two qubits prepared in the state $\ket{\phi^+} = 1/\sqrt{2}(\ket{00} + \ket{11})$, and the later interaction is according to the {transverse-field} Ising model for nearest neighbour interaction between two dipoles~\cite{Ising25} with the following Hamiltonian}
\ba
H = -J \sigma_x\sigma_x -h(\sigma_z \id + \id \sigma_z), 
\ea
where $J$ is the coupling strength with which the dipoles are aligned along the $x$ direction and $h$ is the strength of the external magnetic field along the $z$ direction. The operator that describes the evolution of system and environment is $T_1 = U(J,h,t) = e^{-iH(J, h)t}$ acting on $A_OE_1$. 

The process matrix, as we obtain in {Appendix A4}, is
\ba
W^{A_IA_OB_I}(J,h,t) = \Tr_{E_2} [[U(J,h,t)]],
\ea
where, for a linear operator $x$,
we define $[[x]] = \id\otimes x \KetBra{\id}{\id}\id\otimes x^\dag$.

To detect quantum memory for this process, we start with the PPT criterion~\cite{Peres:1996aa}. As discussed above, we can simply check if the partial transpose of the corresponding state $W^{A_I|A_OB_I} = \rho^{A_I|A_OB_I} = \rho^{AB}$  is {not positive semidefinite, that is if} $\rho^{T_A}\ngeq0$. If so, the state is entangled and an entanglement witness can be constructed by taking the projector of the apppropriate eigenvector of $\rho^{T_{A}}$, as discussed above.
Although the problem is a simple check of negativity of the partial transpose $\rho^{T_{A}}$, it can be cast as a SemiDefinite Program (SDP). 

{An SDP is a particular type of convex optimisation problem and includes the optimisation of a linear function subject to linear constraints. We provide a short introduction to SDPs in Appendix A5.} The problem of detecting entanglement in $\rho$ using the PPT criterion, it can be cast as the following SDP
\ba
\texttt{variable}~~&&Z\ \  \texttt{hermitian semidefinite} \nonumber \\
\texttt{minimize}~~&&\Tr(Z \rho^{T_{A}}) \nonumber \\
\texttt{subject to}~~&& \Tr Z == 1.
\ea
For a negative value of the minimising quantity (over the hermitian semidefinite $Z$), the quantum memory witness is $Z^{T_A}$. 
Although this way of finding a witness is slower than to simply check if $\rho^{T_{A_I}}\ngeq0$, the advantage of the SDP is that we can add restrictions to the witness search, as long as they are linear constraints. 

For example, one might wish to obtain a witness of the form ${Z} = \sum_{i,j}\psi^{A_I}\otimes\psi^{A_O}\otimes E_j^{B_I}$, where Alison performs projective measurements and re-prepares the measured state, while Ben is measuring with some projector $E_j$. In this case the corresponding linear constraint we need to add on the SDP is $PZP = Z$, where $P$ is an operator that swaps the systems $A_I$ and $A_O$ on the witness $Z$.


The second step of the hierarchy developed by Doherty \etal~\cite{Doherty:2004aa} provides stronger criteria for separability. The main idea is that the PPT criterion is applied on a symmetric extension of the state, i.e. on $W^{A_IA_OB_IA_I}$. We present the associated SDPs in Appendix A5. 


We run our code using the PPT criterion across the parameter space $(J,h)$ for fixed $t$. As $t$ is the time of interaction between the system and the memory, for time $t=0$ the process is trivially Markovian for every nonzero $(J,h)$. This is also the case for $J=0$ for every nonzero $(t,h)$ because the magnetic field is affecting both systems separately. In Appendix A6 and A7 we find all the other parameter values for which the process is Markovian and we prove that for $h=0$ and every nonzero $(J,t)$ the process is with classical memory (or Markovian). 

For a fixed time $t=1$ we run the code for $(J, h) \in [0,10]$ and present our findings in Figure~\ref{fig:trace}. The value $\Tr(ZW)$ becomes negative when the PPT criterion has detected a quantum memory while in all other areas, a value close to zero is in general inconclusive. For this reason, we run the SDP testing the second step of the hierarchy. We find the same results, suggesting that the areas of not detecting a quantum memory were probably areas with classical memory. However, as we present in Appendix A7, the points in the Figure where the lines are crossing correspond to Markovian processes. Additionally, we verify that these are the only Markovian points by calculating the measure of non-Markovianity introduced in~\cite{Pollock:2018ab}, which is the {relative entropy~\cite{Vedral:2002aa} between} each process with an associated Markovian one~\cite{Giarmatzi:2018aa}. This measure becomes zero when the process is Markovian. 

\bef
\center
\includegraphics[width=.9\linewidth]{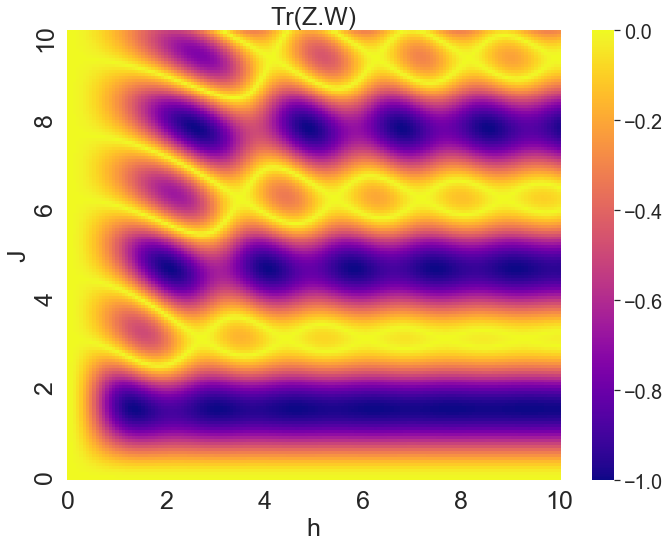}
\caption[]{Heatmap of the value $\Tr(ZW)$ across the parameter space $(J,h) \in [0,10]$ with steps of $1/15$, for $t=1$, where $Z$ is the witness obtained through the PPT criterion and $W$ is the process matrix of our example}
\label{fig:trace}
\eef

\section{Conclusion}
{We provide a novel approach to the study of non-Markovianity that captures genuine quantum multi-time correlations within a process and allows us to detect an environment with quantum memory without process tomography.} Our method expands considerably the investigation of open quantum systems. Our definition of non-Markovianity with classical memory, which leads to mapping entanglement to a quantum memory, provides an experimentally accessible tool and paves the way for further research. It can be used in experimental situations to detect that an external quantum system (environment) retains quantum correlations over time---whether this is desired or not. Furthermore, entanglement leads to many properties for states, which can be mapped to interesting properties for processes, for example multipartite entanglement could correspond to novel types of quantum non-Markovianity in multi-time-step processes. Hence, investigating other properties of entanglement could unveil more information about quantum memory in open quantum systems.

\begin{acknowledgments} We thank Andrew Doherty for significant clarification on their paper, on which our SDPs are based. We thank Gerard Milburn, Kavan Modi, Felix Pollock, and Philip Taranto for discussions. This work was supported by the Australian Research Council (ARC) Centre of Excellence for Quantum Engineered Systems grant (CE 110001013). C.G. is the recipient of a Sydney Quantum Academy Postdoctoral Fellowship. F.C. acknowledges support through an Australian Research Council Discovery Early Career Researcher Award (DE170100712). Furthermore, this publication was made possible through the support of a grant from the John Templeton Foundation. The opinions expressed in this publication are those of the authors and do not necessarily reflect the views of the John Templeton Foundation. Finally, we acknowledge the traditional owners of the land on which the University of Queensland and the University of Technology Sydney are situated, the Turrbal and Jagera people, and the Gadigal of the Eora Nation.
\end{acknowledgments}

%

\end{bibunit}

%
\begin{bibunit}[apsrev4-1_modified]
\appendix
\section{Appendix}
\label{sec:Sup_material}
\setcounter{equation}{0}
\setcounter{figure}{0}
\subsection{Choi representation of processes with classical memory}
{In this section we provide the definition of a process matrix with classical memory and prove that, as a necessary condition, it is proportional to a separable state.}

We consider a finite number $t_1,\dots,t_n$ of time steps. Each time step is associated with an input and an output space, $A^j_I\equiv \mathcal{L}(\mathcal{H}^{A^j_I})$, $A^j_O\equiv \mathcal{L}(\mathcal{H}^{A^j_O})$, $j=1,\dots,n$, corresponding to the input and output of an operation that can be performed on the system at time $t_j$. We use the short-hand $A^j\equiv A^j_I\otimes A^j_O$. We take the last output space to be trivial, so that $A^n\equiv A^n_I$. (Equivalently, we could keep a non-trivial final output space $A^n_O$, but the constraints defining a causally ordered process---or comb---imply that $W^{A^1\dots A^n_I}\otimes\id^{A^n_O}$~\cite{chiribella09b, Oreshkov12}, which means $A^n_O$ is not relevant for calculating observable probabilities.)

{In a} quantum process with classical memory, the environment can acquire classical information about the system, which then feeds forwards to influence the system's evolution at later times.  The acquisition of classical information by the environment can be understood as a measurement: during {the evolution between times} $t_j$ and $t_{j+1}$, the environment {measures the system and obtains a classical} outcome $a_j$. {The conditional evolution of the system, in general, depends on $a_j$. Furthermore, the evolution can depend on the information stored by the environment up to that point, which we represent as a classical variable $x_j$. The most general evolution of this type is represented by a CP map $\mathcal{T}^j_{a_j|x_j}:{A^j_O \rightarrow A^{j+1}_I}$.}

{If the outcome $a_j$ is discarded, the \emph{unconditional} evolution of the system between times $t_j$ and $t_{j+1}$ is represented by the sum $\sum_{a_j}\mathcal{T}^j_{a_j|x_j}$, which must be a CPTP map. In other words, each set of CP maps $\left\{\mathcal{T}^{j}_{a_j|x_j}\right\}_{a_j}$ has to be an instrument.} In Choi represenation, an instrument is given by a set of matrices that satisfy
\begin{equation}
T^{A^j_O A^{j+1}_I}_{a_j|x_j}\geq 0, \qquad \tr_{A^{j+1}_I} \sum_{a_j} T^{A^j_O A^{j+1}_I}_{a_j|x_j} = \id^{A^j_O}.
\end{equation} 

The value of $x_j$ at time $t_j$ can depend on the value of $x$ at all previous times, as well as on all the information hitherto extracted from the system, through an arbitrary conditional probability $P(x_j|\vec{a}_{|j},\vec{x}_{|j})$, where we use the short-hand notation $\vec{a}_{|j}\equiv (a_0,\dots,a_{j-1})$ (and similarly for $\vec{x}_{|j}$). All the classical information in the environment is eventually discarded, leading to the process matrix
\begin{equation}
W^{A^1\dots A^n}_{\textrm{Cl}} = \sum_{\vec{a}\,\vec{x}}\bigotimes_{j=0}^{n-1} T^{A^j_O A^{j+1}_I}_{a_j|x_j}P(x_j|\vec{a}_{|j},\vec{x}_{|j}),
\label{classicalmemory}
\end{equation}
with $\vec{a}\equiv (a_0,\dots a_{n-1})$, $\vec{x}\equiv (x_0,\dots x_{n-1})$.
Note that, since we start with an input space, $A^1_I$, the ``zeroth'' instrument reduces to a set of sub-normalised states, $\left\{T^{A^{1}_I}_{a_0|x_0}\right\}_{a_0}\equiv \left\{\rho^{A^{1}_I}_{a_0|x_0}\right\}_{a_0}$, summing up to a normalised state, $\tr \sum_{a_0} \rho_{a_0|x_0} = 1$. This describes an environment that, given the variable $x_0$, prepares {the} state $\rho_{a_0|x_0}/\tr \rho_{a_0|x_0}$ with probability $P(a_0|x_0)=\tr\rho_{a_0|x_0}$.

{It can be helpful to give a temporal description of the process \eqref{classicalmemory}. Before time $t_1$, the environment contains some classical variable $x_0$. Upon measuring the system, the environment obtains a classical outcome $a_0$. This leaves the system in the (non-normalised) state $\rho_{a_0|x_0}$. At time $t_1$, the environment's information is updated to the classical variable $x_1$, which depends on the previous variables through the conditional probability $P(x_1|a_0,x_0)$. Then, between times $t_1$ and $t_2$, the environment measures the system, obtaining an outcome $a_1$ and resulting in the conditional evolution $\mathcal{T}^1_{a_1|x_1}:{A^1_O \rightarrow A^{2}_I}$ of the system. The memory at $t_2$ updates to $x_2$, through $P(x_2|a_1,a_0,x_1,x_0)$, and so on.}

We can re-define the CP maps to include the conditional probabilities representing the dependency on past information. Thus we set
\begin{align}
T^{A^j_O A^{j+1}_I}_{a_j x_j|\vec{a}_{|j},\vec{x}_{|j}} &:= T^{A^j_O A^{j+1}_I}_{a_j|x_j} P(x_j|\vec{a}_{|j},\vec{x}_{|j}) , \\
\rho^{A^1_I}_{a_0 x_0} &:= \rho^{A^1_I}_{a_0|x_0}P(x_0),
\label{compactinstrument}
\end{align}
where $P(x_0)$ is the marginal probability for the initial variable $x_0$. It is easy to verify that the newly defined maps form properly normalised instruments: 
\begin{align*}
\tr_{A^{j+1}_I} \sum_{a_j x_j} T^{A^j_O A^{j+1}_I}_{a_j x_j|\vec{a}_{|j},\vec{x}_{|j}} &= \id^{A^j_O}, \\
 \tr \sum_{a_0 x_0}\rho_{a_0 x_0} &= 1.
\end{align*}
As now both variables $a_j$, $x_j$ correspond to measurement outcomes in the instrument, we can further simplify the expression by combining the variables as $(a_j, x_j) \mapsto x_j$, $(\vec{a}_{|j}, \vec{x}_{|j}) \mapsto \vec{x}_{|j}$. 
The process matrix now takes the form
\begin{equation}
W^{A^1\dots A^n}_{\textrm{Cl}} = \sum_{\vec{x}} \bigotimes_{j=0}^{n-1} T^{A^j_O A^{j+1}_I}_{x_j|\vec{x}_{|j}}.
\label{classicallong}
\end{equation}
{As this is a particular form of Eq.~\eqref{classicalmemory}, we see that a process matrix has classical memory if and only if it can be written in this way. In other words, we can take Eq.~\eqref{classicallong} as a definition of process with classical memory.}

By relabelling $\vec{x}\mapsto x$, we see that the process matrix \eqref{classicallong} is a particular case of the form
\begin{align} \label{classicalshort}
W^{A^1\dots A^n}_{\textrm{Cl}} &= \sum_{x} \rho^{A^{1}_I}_{x}\otimes T^{A^1_O A^{2}_I}_x\otimes \dots \otimes T^{A^{n-1}_O A^{n}_I}_x, \\ \nonumber
\rho^{A^{1}_I}_{x} &\geq 0,\qquad  T^{A^{j-1}_O A^{j}_I}_x \geq 0,
\end{align}
where we no longer require that the CP maps $T_x$ sum up to a CPTP map. In other words, the process matrix is proportional to a separable state {for the partition $A^1_I|A^1_OA^2_I|\dots|A^{n-1}_OA^n_I$}, as we set out to prove. Note that this is equivalent to a convex combination of product matrices, as easily seen be redefining $\rho_x\mapsto p_x\rho_x$, $p_x=\tr\rho_x$.

We remark that, although all process matrices with classical memory have the form \eqref{classicalshort}, not all matrices of that form are process matrices with classical memory. In other words, it might not be possible to re-write a matrix of the form \eqref{classicalshort} into the form \eqref{classicallong}. This is certainly the case for separable states that do not satisfy the constraints for causally ordered process matrices (combs). However, we conjecture that there are combs of the form \eqref{classicalshort} that cannot be rewritten in the form \eqref{classicallong}, i.e., that cannot be realised as processes with classical memory.

{This conjecture is motivated by the close analogy with ``quantum non-locality without entanglement'' \cite{Bennett1999}. In detail, a multi-time process (or comb) defines a multipartite channel from all output to all input spaces, $\mathcal{W}:\bigotimes_{j}A_O^j\rightarrow \bigotimes_{j}A_I^j$. The process matrix $W$ is the Choi representation of $\mathcal{W}$. {To put this in the context of} the relevant literature, we can take $A^1_I$ to be trivial and $n=3$. Furthermore, assuming matching {dimensions}, we identify $A^{1}_O\equiv A^{2}_I \equiv A$ and $A^{2}_O \equiv A^{3}_I \equiv B$. A process matrix of the separable form \eqref{classicalshort} then corresponds to a channel 
\begin{equation}\label{separablechannel}
\mathcal{W}_{\textrm{sep}} = \sum_x \mathcal{T}_x^{A}\otimes \mathcal{T}_x^{B},
\end{equation}
where each $\mathcal{T}_x^{A}$, $\mathcal{T}_x^{B}$ is a CP (but not necessarily CPTP) map. A channel of the above form is called \emph{separable}, in analogy to separable states \cite{Rains1997}. A subclass is that of LOCC channels, namely channels that can be realised with Local Operations and Classical Communication. For example, an LOCC channel can consist of an instrument $\left\{\mathcal{T}_{a|x}^{A}\right\}_a$ performed on system $A$ and an instrument $\left\{\mathcal{T}_{b|y}^{B}\right\}_b$ on $B$, where $y$ depends probabilistically on $a$ and $x$ (thanks to ``classical communication'' from $A$ to $B$):
\begin{equation}
\mathcal{W}_{\textrm{LOCC}}= \sum_{abxy}P(y|ax)P(x) \mathcal{T}_{a|x}^{A}\otimes \mathcal{T}_{b|y}^{B}.
\label{LOCCchannel}
\end{equation}
More generally, an LOCC channel can involve the composition of multiple instruments on each side, with multiple rounds of communication between $A$ to $B$.
}

{It is relatively simple to see that a channel of the form \eqref{LOCCchannel} corresponds to a process with classical memory, according to our definition. The fact that there exist separable channels that do not have an LOCC realisation \cite{Bennett1999} strongly suggests that there are combs with the separable form \eqref{classicalshort} that cannot be realised as processes with classical memory. The only gap is that combs correspond to a strict subset of multipartite channels. Therefore, it is possible that, when adding the comb constraints, separable channels may fall into the LOCC class, and possibly even in the smaller class of processes with classical memory.}


\subsection{Non-Markovianity witnesses based on dynamical maps}

A large body of work on open quantum systems has been concerned with identifying non-Markovian behaviour in dynamical maps, leading to several witnesses and measures of non-Markovianity \cite{Rivas2014, Breuer16}. Here we compare this approach to our definition of quantum memory and show that common measures of non-Markovianity are not sufficient do decide whether a general process has quantum or classical memory.

A dynamical map is a time-dependent quantum channel, describing the evolution of a system as $\rho(t_1)\mapsto \rho(t)= \mathcal{E}_{t,t_1}\left(\rho(t_1)\right)$. The map $\mathcal{E}_{t,t_1}$ can be seen as the quantum analogue of a conditional probability distribution $P(X_t|X_{t_1})$ for a classical time-dependent stochastic variable $X_t$. Knowledge of these conditional probabilities does not provide full information about the stochastic process, which is defined through joint probabilities $P(X_{t_1},X_{t_2},\dots,X_{t_n})$. Similarly, knowledge of the dynamical map $\mathcal{E}_{t,t_1}$ does not provide full information about the processes matrix $W^{A_1,\dots,A_n}$, which encodes arbitrary multipartite correlations \cite{Pollock:2018ab}. 

Nonetheless, one can use dynamical maps to derive sufficient conditions for non-Markovianity. The most commonly used criteria are based on the notion of \emph{divisibility}~\cite{Wolf2008}. A set of dynamical maps is called divisible if, for any three times $t_1$, $t_2$, $t_3$, the decomposition $\mathcal{E}_{t_3,t_1}=\mathcal{E}_{t_3,t_2}\circ \mathcal{E}_{t_2,t_1}$ holds, where all maps are completely positive. Although not all divisible maps correspond to Markovian processes, all Markovian processes give rise to divisible maps~\cite{Milz2019}. Therefore, a failure of divisibility is a signature of non-Markovianty. Our question is whether it is also a signature of \emph{quantum} memory, as per our definition. Conversely, we also want to know whether a process with classical memory necessarily leads to divisible dynamics. We are going to answer both questions in the negative.

An important observation is that, in the presence of initial correlations between system and environment, dynamical maps are in general not well-defined, as it becomes unclear how to vary the input to the map without specifying the full system-environment state \cite{Shaji2005, Pechukas1994, Stelmachovic2001, Schmid2019}. This already shows that dynamical maps are unsuitable to study memory in two-time-step scenarios (as our in example in the main text) where non-Markovianity can only arise due to system-environment correlations.

Dynamical maps become well defined in scenarios where the first operation is restricted to be a strong state preparation, removing all initial correlations with the environment (a ``causal break'' in the language of Refs.~\cite{Pollock:2018ab, Pollock:2018aa}). In this scenario, we can take the input space associated with the first time-step to be trivial, $A\equiv A_O$. Non-trivial memory effects can arise when we consider two additional time-steps, $B\equiv B_I\otimes B_O$ and $C\equiv C_I$. This leads us to consider process matrices $W^{A_OB_IB_OC_I}$, whose circuit representation is shown in Figure~\ref{fig:supp1}. In such a process, the relevant partition to detect quantum memory is $A_OB_I|B_OC_I$.

\bef
\center
\includegraphics[width=.8\linewidth]{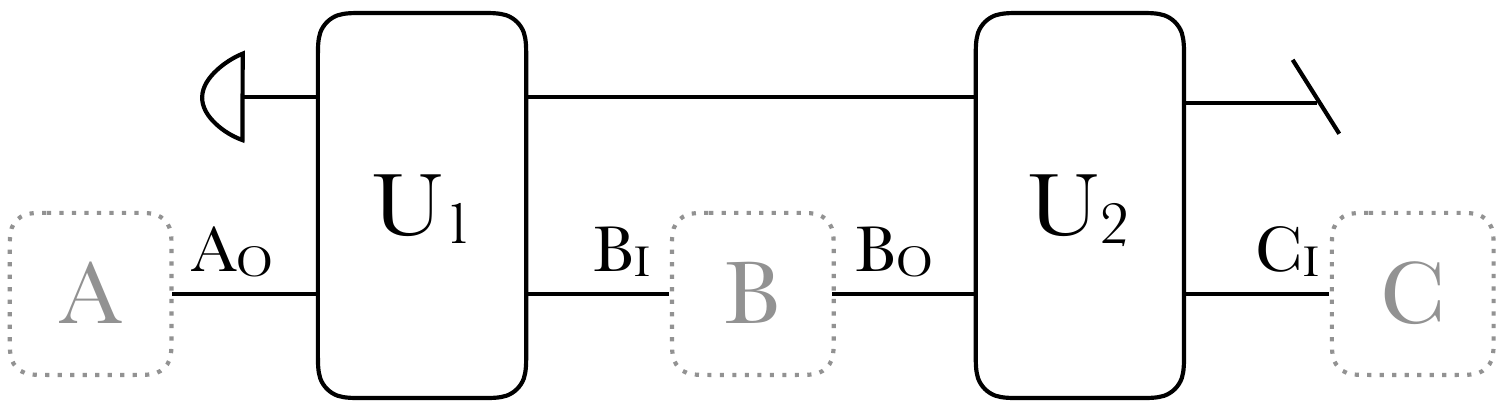}
\caption{Circuit representation of a three-time-step process.}
\label{fig:supp1}
\eef

As this process involves three time-steps, we can associate to it three maps: $\mathcal{E}^{A_0\rightarrow B_I}$ (evolution from $t_1$ to $t_2$), $\mathcal{E}^{B_0\rightarrow C_I}$ ($t_2$ to $t_3$), and $\mathcal{E}^{A_0\rightarrow C_I}$ ($t_1$ to $t_3$). The relation with the process matrix is as follows (see, e.g., Ref.~\cite{costa2016, Milz2020kolmogorovextension}): 
\begin{enumerate}
	\item $\mathcal{E}^{A_0\rightarrow B_I}$ is obtained by discarding the last time-step. It is a general property of a quantum comb (causally ordered process matrix) that tracing out the last input space results in the identity matrix on the last output \cite{chiribella09b}:
	\begin{equation}
	\Tr_{C_I} W^{A_OB_IB_OC_I} = E_1^{A_OB_I}\otimes \id^{B_O}.
	\label{comb4}
	\end{equation}
	$E_1^{A_OB_I}$ is the Choi matrix of the map $\mathcal{E}^{A_0\rightarrow B_I}$, and can be obtained directly from the process matrix through
	\begin{equation}
	E_1^{A_OB_I} = \frac{1}{d^B_O}\Tr_{B_OC_I} W^{A_OB_IB_OC_I},
	\label{recoverE}
	\end{equation}
	where $d^B_O$ is the dimension of $B_O$.
	\item $\mathcal{E}^{A_0\rightarrow C_I}$ is obtained by inserting the identity map as $B$'s operation. 
In Choi representation,
	\begin{multline}
		E_2^{A_OC_I} \\
		= \Tr_{B_IB_O}\left[\left(\id^{A_IC_I}\otimes \left[\left[\id\right]\right]^{B_I B_O}\right) W^{A_OB_IB_OC_I}\right].
	\label{putidentity}
	\end{multline}
	Note that, in expressions such as Eq.~\eqref{putidentity}, a re-ordering of subsystem is implied (i.e., appropriate subsystem swaps should be applied to matrices so that subsystems are all in the same order, e.g., the standard order $A_OB_IB_OC_I$).
	\item $\mathcal{E}^{B_0\rightarrow C_I}$ is not uniquely specified by $W$, as it depends on the state preparation at $A_O$. However, in the following we will not need an explicit expression of this map. (In fact, divisibility criteria typically only require the existence of \emph{any} CP map $\mathcal{E}^{B_0\rightarrow C_I}$, not necessarily operationally defined \cite{Milz2019}).
\end{enumerate}

The first thing we want to prove is that divisibility does not imply classical memory. To this end, we present a process matrix that is entangled across the partition $A_OB_I|B_OC_I$ and whose associated maps satisfy divisibility. The example is 
\begin{align}
W^{A_OB_IB_OC_I} &= \left[\Phi\right]^{A_OB_IC_I}\otimes \id^{B_O},
\label{GHZ} \\
\ket{\Phi} &= \ket{000}+\ket{111},
\end{align}
with the notation $\left[\Phi\right]\equiv \ket{\Phi}\bra{\Phi}$.
This process can be realised through the circuit in Fig.~\ref{fig:supp2}. As the ``GHZ state'' $\ket{\Phi}$ is entangled across any bipartition, it is clear that this process matrix has a quantum memory. A simple calculation reveals the Choi matrices of the relevant maps to be
\begin{align}
E_1^{A_OB_I} &= \tr_{C_I}\left[\Phi\right]^{A_OB_IC_I} = \left(\left[00\right]+\left[11\right]\right)^{A_OB_I}, \\
E_2^{A_OC_I} &= \Tr_{B_I} \left[\Phi\right]^{A_OB_IC_I} = \left(\left[00\right]+\left[11\right]\right)^{A_OC_I}.
\end{align}
Thus, the two maps are equal, each describing a dephasing channel $\rho\mapsto \left[0\right]\bra{0}\rho\ket{0}+\left[1\right]\bra{1}\rho\ket{1}$. These maps are clearly compatible with divisible dynamics. In particular, the criterion in Ref.~\cite{Rivas10} states that evolution is divisible if and only if the probability of success in a one-shot state discrimination task cannot increase. This is clearly true in our example: for any given state preparation at time $t_1$, tomography at $t_2$ would reveal the same state as at time $t_3$. Memory signatures in process \eqref{GHZ} are encoded in genuinely tripartite correlations across $A_OB_IC_I$. Hence, divisibility does not imply classical memory.

\bef
\center
\includegraphics[width=.8\linewidth]{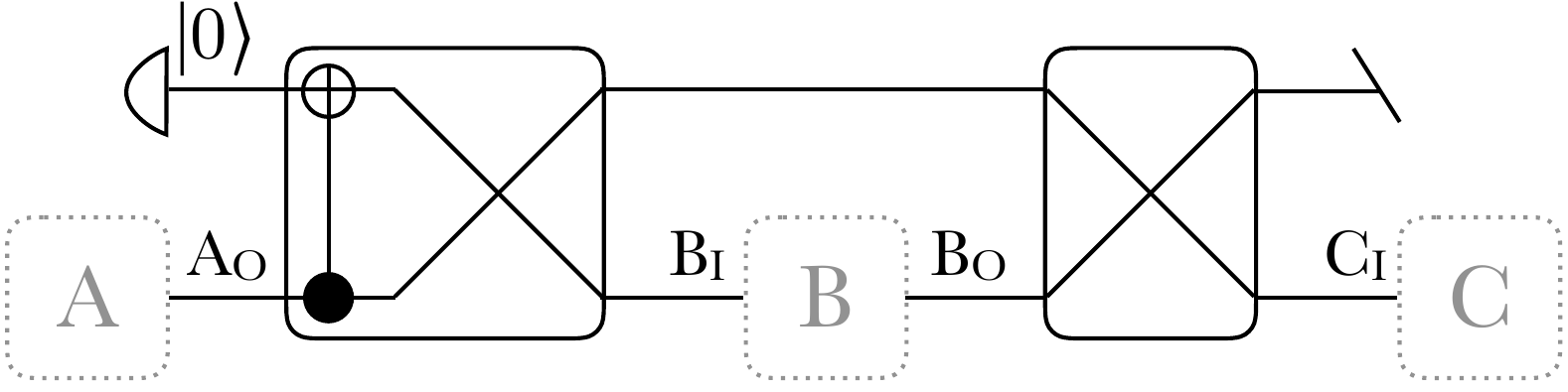}
\caption{Process with quantum memory, identified as Markovian by the dynamical maps approach.}
\label{fig:supp2}
\eef

Let us now show that non-divisibility does not imply quantum memory. We will present a process with classical memory and show that its associated maps are non-divisible. To this end, for divisibility, it is sufficient to consider the criterion proposed in Ref.~\cite{Breuer09}: given two initial states $\rho_0$, $\rho_1$, prepared at $t_1$, their trace distance cannot increase under divisible dynamics; the trace distance is defined as $D(\rho_0,\rho_1):=\frac{1}{2}\left\|\rho_0-
\rho_1\right\|_1$, $\left\|\Delta\right\|_1:=\Tr\sqrt{\Delta^{\dag}\Delta}$. We want to present a process with classical memory that increases the trace distance of two states, {implying non-divisibility}. The example is
\begin{equation}
W^{A_OB_IB_OC_I} =\sum_{a=0}^1 \left[a\right]^{A_O}\otimes \rho^{B_I}\otimes\id^{B_O}\otimes \left[a\right]^{C_I},
\label{Dephased}
\end{equation}
with $\rho$ an arbitrary density matrix. {This process has the form \eqref{classicallong} and thus can be realised with classical memory, through the circuit in Figure~\ref{fig:supp3}. Explicitly, between the first and second time step ($A_O$ to $B_I$) the environment measures the system in the computational basis, yielding outcome $a$, and prepares the unrelated state $\rho$. Between second and third step ($B_O$ to $C_I$), the system's state is discarded and the pure state $\ket{a}$ is prepared, which only requires the environment to keep memory of the classical output $a$ from the first time step. This results in a dephasing channel from the first to the third step ($A_O$ to $C_I$).}
Now, consider states $\rho_0=\left[0\right]$, $\rho_1 = \left[1\right]$ prepared at $t_1$. Their trace distance drops to $0$ at $t_2$, as any state is replaced by $\rho$. However, at $t_3$ the initial value is restored, $D(\left[0\right],\left[1\right])= 1$, as both states are mapped identically from $A_O$ to $C_I$. {Hence, a process with classical memory can involve non-divisible maps, which means that non-divisibility does not imply a quantum memory.}

\bef
\center
\includegraphics[width=.8\linewidth]{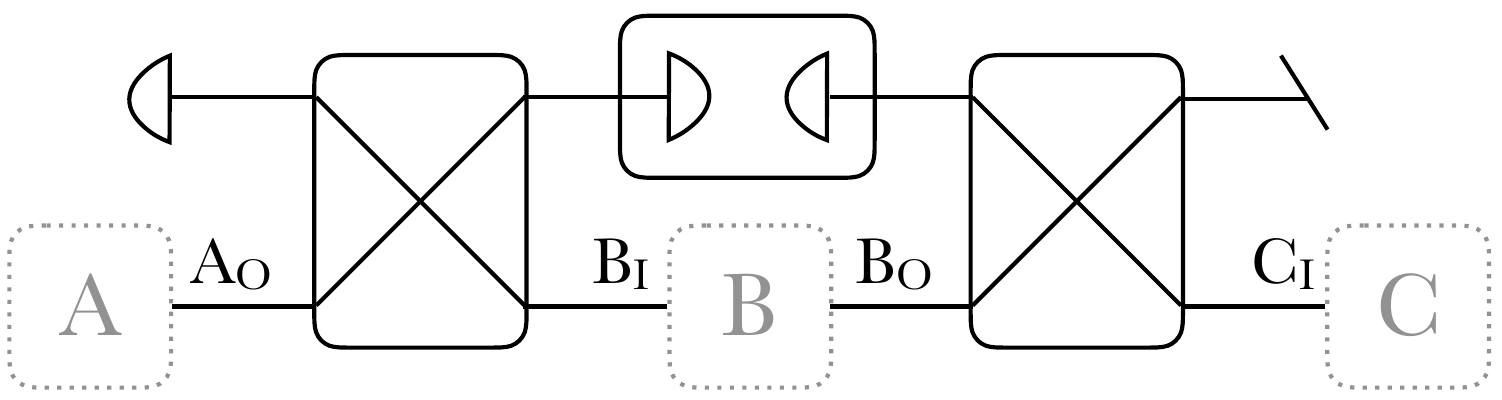}
\caption{Process with classical memory, where the distinguishability between states can increase.}
\label{fig:supp3}
\eef

This  example is not surprising, as it is known that the (classical analogue of the) trace-distance criterion is violated by classical stochastic processes \cite{Chruscinski2011}, and thus it is fully expected that it can be violated by quantum processes with classical memory.

\subsection{Initial system-environment correlations affecting the memory}
In this section we discuss what happens to a general 2-time-step scenario (one system-environment interaction), with initial system-environment correlations, described in the initial joint state $\rho^{E_1A_I}${, where $E_1$ denotes the environment state space at the initial time}.  The process matrix is obtained using the rules to construct quantum combs, by sequential composition of processes~\cite{chiribella09b}. We start with the initial state $\rho^{E_1A_I}$ and the joint unitary in the form of its Choi matrix $T^{E_1 A_OE_2B_I}$, and combining them using the \emph{link product}, defined as 

\begin{multline}
W^{A_IA_OB_I} =\rho*T \nonumber \\ 
 = \tr_{E_1E_2} \left[ \left(\rho^{E_1A_I} \otimes \id^{E_2A_OB_I} \right) 
  \left(T^{E_1^{\textup{T}} A_OE_2B_I}\otimes \id^{A_I}\right) \right],
\end{multline}
where the superscript on the subsystem $E_1$ denotes partial transposition with respect to the subsystem.

\subsubsection{Case 1: $\rho$ is a product state}
In this case $\rho^{E_1A_I} = \rho_1^{E_1}\otimes \rho_2^{A_I}$, with $\rho_1$ and $\rho_2$ being arbitrary states. The process matrix becomes 
\begin{align}
W&^{A_IA_OB_I} =\nonumber \\ 
&\tr_{E_1E_2} \left[ \left( \rho_1^{E_1}\otimes \rho_2^{A_I} \otimes \id^{E_2A_OB_I}\right)\right. \nonumber \\
 &\left.   \left(T^{E_1^T A_OE_2B_I}\otimes \id^{A_I}\right)  \right] \nonumber \\
  &=\rho_2^{A_I} \otimes T_s^{A_OB_I} ,
\end{align}
where $T^{A_OB_I}_s = \tr_{E_1E_2}\left[\left( \rho_1^{E_1}\otimes \id^{A_OE_2B_I}\right) T^{E_1^T A_OE_2B_I}\right]$ is the channel for the system after tracing out the environment. We see that this is the form of a Markovian process matrix. 

\subsubsection{Case 2: $\rho$ is a separable state}
In this case, $\rho^{E_1A_I} = \sum_{j} q_j \rho_{j}^{E_1}\otimes \sigma_{j}^{A_I}$, with $\rho_{j}, \sigma_{j}\geq 0$, $0\leq q_j\leq 1$, and $\sum_j q_j =1$. Now the process matrix becomes
\begin{align}
W&^{A_IA_OB_I} =\nonumber \\ 
  &\tr_{E_1E_2} \left[ \left( \sum_{j} q_{j} \rho_j^{E_1}\otimes \sigma_j^{A_I} \otimes \id^{E_2A_OB_I} \right)\right. \nonumber \\
 &\left.   \left(T^{E_1^T A_OE_2B_I}\otimes \id^{A_I}\right)  \right] \nonumber \\ \label{classical2}
  &=\sum_{j} q_{j} \sigma_j^{A_I} \otimes T_j^{A_OB_I},
\end{align}
where, $\forall j$,
$$T_j^{A_OB_I} = \tr_{E_1E_2}\left[\left( \rho_j^{E_1}\otimes \id^{A_OE_2B_I}\right) T^{E_1^T A_OE_2B_I}\right] $$ 
represents a CPTP map.
We see that, for $j>1$, Eq.~\eqref{classical2} has the form of a process with classical memory, Eq.~\eqref{classicallong}. 

We conclude from the above that for a two-time-step process, and any kind of system-environment interaction, there can be no quantum memory in the environment if there is no initial entanglement between the environment and the system.

\subsection{Process matrix of our example}
For the process matrix for our example we start with the one written above for the general two-time-step process 
\begin{equation}
\rho*T = \tr_{E_I}  \left[ \left(\rho^{{E_1}^TA_I}\otimes \id^{A_OB_IE_2}\right) \left(\id^{A_I}\otimes T \right)\right],
\end{equation}
with the partial transpose acting on the common subsystem that is traced over.
We then substitute $T= \left[\left[U\right]\right]$ and, omitting the tensor product symbols,
\begin{equation}\label{doubleproj}
\left[\left[U\right]\right] = \sum_{jkj'k'}\ketbra{jk}{j'k'}^{A_OE_1}\left(U  \ketbra{jk}{j'k'} U^{\dag}\right)^{B_IE_2},
\end{equation}
while the initial state is $\rho=\proj{\phi^+}$, so that $\rho^{T_{E_1}}=\frac{1}{2}\sum_{il} \ketbra{i}{l}^{A_I}\ketbra{l}{i}^{E_1}$. The link product then gives
\begin{multline} \label{beforetracing}
\proj{\phi^+}*[[U]]\\
=\frac{1}{2}\sum_{ijk}\ketbra{k}{k'}^{A_I}\ketbra{j}{j'}^{A_O}\left(U  \ketbra{jk}{j'k'} U^{\dag}\right)^{B_IE_2},
\end{multline}
which is, up to normalisation, the same expression as \eqref{doubleproj}, with the first two factors swapped and $E_I$ substituted with $A_I$. The process matrix describing the first qubit is obtained by tracing out $E_2$.


\subsection{SDP: Higher-order criteria}

For the second level of the hierarchy, the problem is no longer a simple computational task---we need to find if there exist a symmetric extension $\tilde{\rho}^{ABA}$ of a state $\rho^{AB}$ (corresponding to a process matrix $W^{AB}$), such the state and all its partial transposes with respect to $A$ and $B$ are positive. The latter conditions are linear constraints on $\tilde{\rho}^{ABA}$ and so the problem can be cast as an SemiDefinite Program which can be solved efficiently~\cite{Doherty:2004aa}. We develop this method as it is a stronger test of non-separability, although in our example it turns out that the first level completely characterizes the type of non-Markovinity.

{An SDP has a primal and a dual form. The primal form is written as
\ba
\texttt{variable}~~&&\bf{x} \nonumber \\
\texttt{minimize}~~&&c^T\bf{x} \nonumber \\
\texttt{subject to}~~&& F(\bf{x}) \geq 0,
\ea
where c is a known vector, ${\bf x} = (x_1, \cdots, x_n)$ and $F({\bf x} ) = F_0 + \sum_i x_i F_i$, for some fixed hermitian matrices $F_i$, is a semidefinite matrix. In the case where $c=0$ there is no function to minimise and the problem reduces to whether the constraints are satisfied for some value of ${\bf x}$.} This is called the \emph{feasibility problem}. {Its associated dual form is written as 
\ba
\texttt{variable}~~&& Z\ \ \texttt{hermitian}  \nonumber \\
\texttt{maximize}~~&& -\Tr (F_0 Z) \nonumber \\
\texttt{subject to}~~&& Z \geq 0, \nonumber \\
&& \Tr (F_i Z) = c_i
\ea
}

In our case, the feasibility problem is formulated as follows
\ba
\texttt{variable}~~&&\tilde{\rho}~~\texttt{semidefinite} \nonumber \\
\texttt{minimize}~~&&0 \nonumber \\
\texttt{subject to}~~&&\tilde{\rho} \oplus \tilde{\rho}^{T_A} \oplus \tilde{\rho}^{T_B} \geq 0 \nonumber \\
&& \Tr_{A'} \tilde{\rho} = \rho \nonumber \\ 
&& \texttt{swap$_{A,A'}$}(\tilde{\rho}) = \tilde{\rho},
\ea
{where we have omitted the spaces the states live on, $\rho^{AB}$ and $\tilde{\rho}^{ABA'}$. This SDP tries to find an extension of ${\rho}$, $\tilde{\rho}$, such that it is separable according to the PPT criterion. $\tilde{\rho}$ is constrained to be an extension of $\rho$  with the last two constraints and that it is a positive semidefinite matrix.} Note that a block diagonal matrix is non-negative if and only if each of its block matrices is non-negative.

The primal SDP detects entanglement when its output is that the problem is infeasible, {i.e. it cannot find an extension of $\rho$ such that it is separable according to the PPT criterion}. The dual SDP {minimises a value $\Tr(Z \tilde{\rho})$ to find a witness for $\tilde\rho$; from which we find a witness for $\rho$}.  To explicitly write the dual form we need some definitions. Due to symmetries on the extension $\tilde{\rho}$, it can be written as follows 
\ba
\trho = G(x) = G_0 + \sum_Jx_JG_J
\ea
where $G_0$ is a fixed part that depends only on the given $\rho$ and $x_J$ is the coefficients for the rest of the basis elements $G_J$. In particular, $G_0 =\Lambda(\rho)$, and $\Lambda$ is a linear map from $\cal{H}^A\otimes\cal{H}^B$ to $\cal{H}^A\otimes\cal{H}^B\otimes \cal{H}^A$. {If we then define $F(x) = \tilde{\rho} \oplus \tilde{\rho}^{T_A} \oplus \tilde{\rho}^{T_B}$ we get that $F_0 = G_0\oplus G_0^{T_A}\oplus G_0^{T_B}$ and $F_i = G_J\oplus G_J^{T_A}\oplus G_J^{T_B}$.}

The dual form of the SDP then writes 
\ba
\texttt{variable}~~&&Z~~\texttt{hermitian semidefinite} \nonumber \\
\texttt{maximize}~~&&-\Tr(F_0Z)\nonumber\\
\texttt{subject to}~~&&Z \geq 0 \nonumber\\
&&\Tr(F_iZ)=0,
\ea
For a positive value of $-\Tr(F_0Z)$, which yields a negative one for $\Tr(F_0Z)$, we can obtain a witness $\tilde{Z}$ for $\rho$ through $\tilde{Z} = \Lambda^*(Z)$, where $\Lambda^*$ is the adjoint map of $\Lambda$. {The codes for the feasibility and dual problem for our example are publicly available on GitHub~\cite{sdps}}.


\subsection{Our example: classical memory in the $h=0$ case}
In this case the Hamiltonian is
\begin{equation}
H(J)=-J\sigma_x\otimes\sigma_x,
\end{equation}
so it has product eigenstates, built from the Pauli matrix's eigenstates $\sigma_x\ket{\mu} =\mu\ket{\mu}$, $\mu=\pm 1$:
\begin{align}
H(J)&\ket{\mu}\ket{\nu} =-J \mu\nu \ket{\mu}\ket{\nu}, \\
U(J,t)&\ket{\mu}\ket{\nu} =e^{-J \mu\nu t} \ket{\mu}\ket{\nu}, \\\nonumber
\mu, \nu& = \pm 1.
\end{align}

Using the eigenstates of $H(J)$ in the definition of the Choi isomorphism, and taking the partial trace over $E_O$ of expression \eqref{beforetracing}, we obtain for the process matrix
\begin{align*}
W&^{A_IA_OB_I}(J,t)\\
=&\frac{1}{2}\sum_{\substack{\mu_1\mu_2 \\ \nu_1\nu2}}
\ketbra{\nu_1 \mu_1 }{\nu_2\mu_2}^{A_IA_O}\tr_{E_O}\left(U  \ketbra{\mu_1 \nu_1}{\mu_2 \nu_2} U^{\dag}\right)^{B_IE_O} \\
=&\frac{1}{2}\sum_{\mu_1\mu_2\nu} e^{i J (\mu_1- \mu_2)\nu t}
\ketbra{\nu}{\nu}^{A_I}\ketbra{\mu_1}{\mu_2}^{A_O}\ketbra{\mu_1}{\mu_2}^{B_I}.
\end{align*}
We see that this matrix is an equal mixture of two product matrices:
\begin{align*}
W&^{A_IA_OB_I}(J,t)= \frac{1}{2}\sum_{\nu=\pm 1} 
\ketbra{\nu}{\nu}^{A_I}[[e^{ i \nu J \sigma_x  t}]]^{A_O B_I}.
\end{align*}
Therefore, the process we are describing is the equal mixture of two Markovian processes: for $\nu=\pm 1$ respectively, these correspond to processes where the system is initially in state $\ket{\nu}$, while from $A$ to $B$ it evolves according to the (single-qubit) Hamiltonian $H_{\nu}= -\nu J  \sigma_x $. Thus, even though in our model the initial system-environment state is entangled, the process for the system alone can be reproduced by substituting the environment with a classical variable, which samples with equal probabilities the values $\nu=\pm 1$, and determines both the initial state and the evolution from first to second time-step.

\subsection{Our example: markovian points}
Here we find the parameters $J$, $h$, for which the process is Markovian. A sufficient condition is that the unitary matrix that represents the system-environment evolution factorises: $U(J,h,t)= e^{-iH(J, h)t} = U_s\otimes U_e$. As $U(J,h,t)=U(J t, h t, 1)$, it is sufficient to consider the $t=1$ case. 

We find that the unitary factorises for the parameter values
\begin{align*} \label{discreteJ}
J &= \pi k_1,\quad k_1\in \mathbb{Z},\\
h &= \pm\frac{\pi}{2}\sqrt{k_2^2-k_1^2}, \quad k_2\in \mathbb{Z}, \quad k_2^2\geq k_1^2,
\end{align*}
plus an additional point $h=0, J=\pi/2$, for which $U(J,h,1)=i\sigma_x\otimes\sigma_x$.

In principle, there could be parameter values for which the evolution is not trivial but the process is Markovian. We verified that this is not the case numerically: for a range of parameters, we calculated the {relative entropy~\cite{Pollock:2018ab,Vedral:2002aa} between} $W$ and the corresponding Markovian process $W_M = \rho^{A_I}\otimes T^{A_OB_I}$, where $\rho^{A_I}$ and $T^{A_OB_I}$ are obtained as partial traces of $W$. The relative entropy between two states $\sigma$ and $\rho$ is $ S(\sigma||\rho) = \Tr \sigma(\ln \sigma - \ln \rho)$. Note that our states corresponding to process matrices are non-normalised, so we used $\sigma = W/(\Tr W)$ and $\rho = W_M/(\Tr W_M)$.

We found that the relative entropy vanishes (and hence the process is Markovian) only for the parameter values found above. The values of the relative entropy are shown in Figure~\ref{fig:distance}.
\bef
\center
\includegraphics[width=.9\linewidth]{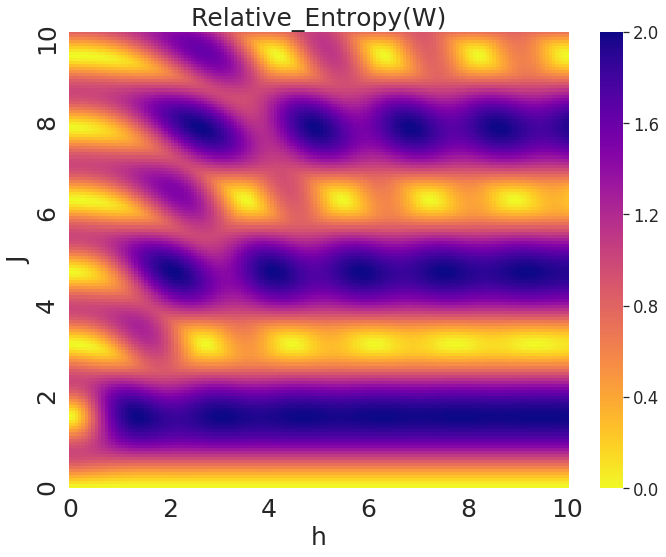}
\caption{Relative entropy $S(W||W_M)$ across the parameter space $(J,h) \in [0,10]$ for $t=1$.}
\label{fig:distance}
\eef


%

\end{bibunit}

\end{document}